\documentclass[useAMS,usenatbib]{mn2e}

\DeclareSymbolFont{cmletters}{OML}{cmm}{m}{it}
\DeclareMathSymbol{v}{\mathalpha}{cmletters}{"76}

\usepackage{amsmath}
\usepackage{amssymb}
\usepackage{epsfig}
\usepackage{graphicx}
\usepackage{ifthen}
\usepackage{latexsym}
\usepackage{rotating}
\usepackage{subfigure}
\usepackage{times,epsf}
\usepackage{txfonts}
\usepackage{varioref}
\usepackage{verbatim}
\usepackage{url}
\usepackage{color}
\usepackage[dvipsnames]{xcolor}
\usepackage[T1]{fontenc}
\usepackage{float}

\newcommand{\be}{\begin{equation}}
\newcommand{\ee}{\end{equation}}
\newcommand{\bea}{\begin{eqnarray}}
\newcommand{\eea}{\end{eqnarray}}

\newcommand\aapr{Astron. \& Astrophys. Rev.}

\newcommand\apj{Astrophysical Journal}
\newcommand\apjl{Astrophysical Journal Letters}

\newcommand\aap{Astronomy \& Astrophysics}

\newcommand\nat{Nature}

\newcommand\mnras{Monthly Notices of the Royal Astronomical Society}

\newcommand\ARAA{Ann. Rev. Astron. Astrophys.}

\newcommand\araa{\ARAA}

  \definecolor{gray}{rgb}{0.6,0.6,0.6}
  \definecolor{green}{rgb}{0,0.6,0}

  \newcommand{\Msun}{M_{\odot}}

\title[QPOs from Oscillating, Precessing Hot, Thick Flow]{High-Frequency and Type-C QPOs from Oscillating, Precessing Hot, Thick Flow}
\author[P. C. Fragile, O. Straub \& O. Blaes]
       {P. Chris Fragile$^1$\thanks{E-mail: fragilep@cofc.edu},
	Odele Straub$^{2,3}$, and Omer Blaes$^{4}$\\
        $^1$ Department of Physics and Astronomy, College of Charleston, Charleston, SC 29424, USA\\
        $^2$ 
             LUTH, Observatoire de Paris, CNRS UMR 8102, Universit\'e Paris-Diderot, 5 place Jules Janssen, 92195 Meudon, France\\
        $^3$ 
             Institute of Physics, Faculty of Philosophy and Science, Silesian University in Opava, 13 Bezru\v covo n\'am., 746 01 Opava, Czech Republic \\
        $^4$ Department of Physics, University of California, Santa Barbara, CA 93106, USA}

\begin{document}

\maketitle

\label{firstpage}

\begin{abstract}
Motivated by recent studies showing an apparent correlation between the high-frequency quasi-periodic oscillations (QPOs) and the low-frequency, type-C QPO in low-mass, black hole X-ray binaries (LMXBs), we explore a model that explains all three QPOs in terms of an oscillating, precessing hot flow in the truncated-disk geometry.  Our model favors attributing the two high-frequency QPOs, often occurring in a near 3:2 frequency ratio, to the breathing and vertical epicyclic frequency modes of the hot, thick flow, although we can not rule out the Keplerian and $m=-1$ radial epicyclic modes.  In either case, the type-C QPO is attributed to precession.  The correlation of the QPOs comes from the fact that all three frequencies are associated with the same geometrical structure.  While the exact QPO frequencies are sensitive to the black hole mass and spin, their evolution over the course of an outburst is mainly tied to the truncation radius between the geometrically thin, optically thick disk and the inner, hot flow.  We show that, in the case of the LMXB GRO J1655-40, this model can explain the one simultaneous observation of all three QPOs and that an extrapolation of the model appears to match lower frequency observations where only two of the three components are seen.  Thus, this model may be able to unify multiple QPO observations using the properties of a single, simple, geometrical model.
\end{abstract}


\section{Introduction}
\label{sec:introduction}

\citet{Motta14a}, and before them \citet[][hereafter PBK]{Psaltis99}, have shown an apparent correlation between high ($>100~\text{Hz}$) and low ($\lesssim10~\text{Hz}$) frequency quasi-periodic oscillations (QPOs) in the power density spectra of low-mass X-ray binaries (LMXBs).  Such correlations have most frequently been explained using a relativistic precession model \citep{Stella98, Stella99}.  One common criticism of such models is their reliance on test particle frequencies, which are not expected to be relevant in an extended accretion flow.

Many other models for high- and low-frequency QPOs have been proposed, though they usually do not simultaneously account for both types of QPOs.  Global oscillation modes, for one, have been shown to be successful at explaining high-frequency QPOs (HFQPOs), especially those occurring in near 3:2 frequency ratios \citep{Rezzolla03, Torok15}.  [\citet{Abramowicz01} were the first to notice that multiple black hole X-ray binaries exhibited such 3:2 frequency ratio pairs of HFQPOs \citep[see also, e.g.,][]{McClintock06,Remillard06}.]  Of particular note, \citet[][hereafter BAF]{Blaes06} demonstrated that the so-called breathing and vertical epicyclic modes always occur in a near 3:2 frequency ratio for a relativistic gas, regardless of the black hole spin or the angular momentum distribution.  However, as we show in this paper, the BAF result only holds in the slender-torus limit.

Low-frequency QPOs have generally received less attention.  One model that has been especially successful in explaining them is the Lense-Thirring precession model of \citet[][hereafter IDF]{Ingram09}. This model differs from the relativistic precession model of \citet{Stella98} in that the precessing object is not a test particle, but an entire geometrically thick flow. (\citealt{Schnittman06} also considered a precessing ring.)  The IDF model assumes a truncated disk geometry, such as is often used to fit observations of the low/hard state of black hole X-ray binaries \citep{Done07}.  In this picture, the accretion flow is represented by a standard thin disk that is truncated well outside the innermost stable circular orbit (ISCO), with the interior region filled by a geometrically thick, hot flow \citep{Esin01}.  The QPO is attributed to the Lense-Thirring precession of this hot, thick flow.  Further, it is the location of the truncation radius in this model that sets the frequency of the QPO by setting the outer geometrical extent of the precessing flow.  During a given outburst, this truncation radius moves in, forcing the precession, and hence QPO frequency, to increase.  The QPO ultimately disappears at a frequency $\nu \sim10~\text{Hz}$, as at this frequency, the truncation radius has nearly reached the ISCO, and the source transitions into the disk-dominated, high/soft state.  The phenomenology of this model matches well that of the type-C QPO seen in many black hole systems \citep[e.g][]{Homan12,Homan15,Motta15}. 

In this paper, we combine global oscillation, high-frequency models with the IDF low-frequency model by presuming that it is the same geometric structure that is both oscillating and precessing to produce the observed spectrum of QPOs.  The correlations between the high- and low-frequency QPOs seen by \citet{Motta14a} are then naturally explained by the fact that all the QPOs are being produced by the same geometric structure.  As we will show in this paper, the properties of the hot, thick flow set the QPO frequencies (for a given black hole mass and spin) and also dictate how the QPO frequencies should evolve during an outburst, based on the location of the truncation radius.  In this paper, we concentrate on fitting the observations of simultaneous QPOs in the LMXB GRO J1655-40.

\section{Mode Frequencies of Non-Slender Tori}
\label{sec:tori}

In this section, we present the oscillation modes of non-slender, hydrodynamic, non-self-gravitating, constant angular momentum tori around a Kerr black hole.  We use these tori as a proxy for the hot, thick flow in the truncated disk model.  The theory of toroidal accretion structures is based on Boyer's condition that the boundary of any barotropic, stationary, perfect fluid body has to be a constant pressure and constant density surface \citep{Abramowicz78}.  Such tori are differentially rotating and known to be dynamically unstable to global, non-axisymmetric perturbations, the so-called Papaloizou-Pringle instability \citep[PPI,][]{Papaloizou84}.  They would likely also develop turbulent stresses due to the magneto-rotational instability (MRI) in the presence of a weak magnetic field \citep{Balbus98}.  However, as stated, we only use the torus as a proxy for the hot, thick part of the accretion flow.  Global MRI simulations of radiatively inefficient accretion flows \citep[e.g.][]{Narayan12} produce flow structures which can be quite geometrically thick and with sub-Keplerian specific angular momentum profiles.  These features are indicative of substantial vertical and radial pressure support, just as in the torus structures that we are considering here.  Actual torus structures with pressure maxima are also seen in global MHD simulations of accretion flows \citep[e.g.][]{DeVilliers03}.  Moreover, simulations of MRI turbulence in stratified shearing boxes \citep[e.g.][]{Hirose09,Blaes11} have already shown that the oscillation modes we study are present, so there is no reason to think that turbulence would prevent them from being present in the real accretion systems we are modeling.  Probably the most serious issue is that the frequencies of some of our modes will be sensitive to our assumed constant specific angular momentum distribution.  Such an angular momentum distribution, which we have assumed for simplicity, is generally not found in global MRI simulations.  We comment on this further below.

\subsection{Oscillation modes}
\label{sec:modes} 

\begin{figure*}  
   \includegraphics[width =0.96\textwidth]{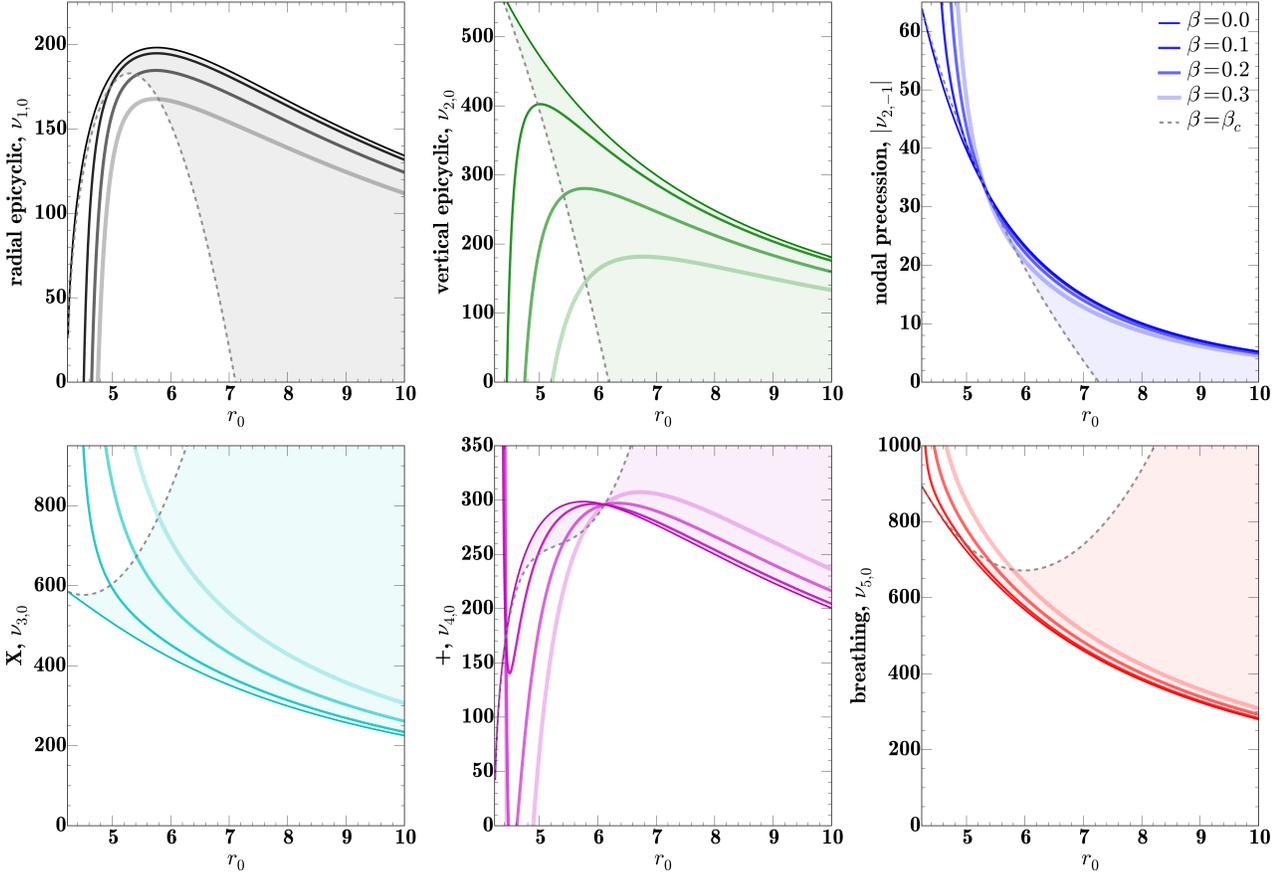}
   \caption{The frequency behavior of the lowest order oscillation modes of a torus.  We use $M_{\rm BH} = 5.4 \Msun$, $a_* = 0.5$, and $n = 3$.  The {\it top row} shows the axisymmetric radial and vertical epicyclic modes, as well as the nodal precession mode, and is a reproduction of the results found by \citet{Straub09}.  The {\it bottom row} shows the X, plus, and breathing modes.  The frequencies of thicker tori are depicted with thicker (and fainter) lines.  Only solutions with $\beta \le \beta_c$ (shaded regions) are considered meaningful.  
}
 \label{fig:frequencies}
\end{figure*}

\citet{Straub09} derived fully relativistic formulae for the frequencies of the five lowest order oscillation modes of non-slender, constant specific angular momentum tori in terms of: the black hole mass $M_{\rm BH}$ and dimensionless spin $a_*$; the torus thickness parameter $\beta$; the radial position of the pressure maximum $r_0$; the polytropic index $n$; and the azimuthal wave number $m$.  They studied in detail the properties of the two epicyclic modes and found that the axisymmetric($m=0$) radial ($i=1$) and vertical ($i=2$) epicyclic frequencies decrease when the torus gets thicker.  We extend their work and calculate the pressure corrections to the other oscillation modes\footnote{We have also corrected some errors in the calculation in \citet{Straub09}.  In particular, we derive the second order pressure corrections from the exact form of the enthalpy function.  We also corrected an error in the relativistic
Papaloizou-Pringle equation of \citet{Abramowicz06}.  (This latter error vanishes in the slender torus limit, and therefore does not affect the results of that paper.)}; 
these are the X ($i=3$), plus ($i=4$), and breathing ($i=5$) modes. The non-slender torus frequency  
\be 
  \nu_{i, m} = \left(\bar\omega_i^{(0)} + m + \beta^2 \bar\omega_i^{(2)}\right) \nu_{\rm K}
  \label{eq:nu_i}
\ee 
of any oscillation mode $i$, is composed of the slender torus (or test particle)  frequency $\bar\omega_i^{(0)}/(2 \pi)$ \citep[calculated in][]{Blaes07} and the second order pressure correction $\bar\omega_i^{(2)}$ 
The Keplerian frequency is simply given by $\nu_{\rm K} = \Omega_{\rm K} c^3/(2 \pi G M_{\rm BH})$, where $c$ is the speed of light, $\Omega_{\rm K}$ the Keplerian angular velocity, and $G$ the gravitational constant.  Intuitively, one may guess that the frequencies of all oscillation modes would decrease with torus thickness.  This is, however, not the case for the acoustic modes, whose frequencies are roughly proportional to $c_{s}/H$, which is roughly constant, where $c_{s}$ is the sound speed and $H$ is the scale height. We show the behavior of the X, plus, and breathing modes of non-slender tori (for $a_*=0.5$) in Figure~\ref{fig:frequencies}.  The largest possible surface of a torus is given by the marginally closed surface, which exhibits a cusp at its inner edge; this surface defines the critical torus thickness, $\beta_c$.  In Figure~\ref{fig:frequencies}, we truncate our solutions whenever this limit is reached. 
Furthermore, since the non-slender formulae were obtained by means of a second order perturbative calculation in torus thickness, we restrict ourselves to only consider solutions with $\beta \leq 0.3$.

We emphasize that these mode frequencies have been calculated for tori with constant specific angular momentum.  At least in the slender torus limit, the
epicyclic mode frequencies are completely independent of the internal properties
of the torus, including its specific angular momentum distribution
\citep{Abramowicz06}.  In addition, \citet{Blaes07} have shown that the
slender torus breathing mode frequency
depends only weakly on the specific angular momentum distribution.

\subsection{Precession mode}
\label{sec:LT} 

Along with oscillating, the hot, thick flow (or torus) may also precess.  The lowest order precession mode is given by the $m=-1$ vertical epicyclic ($i=2$) mode, that is $\nu_{\rm prec} = |\nu_{2, -1}|$.  Far from the black hole, this frequency approaches the Lense-Thirring precession frequency for a tilted flow. 

In our model, as in IDF, we assume this precession frequency manifests itself in the power density spectrum as the type-C QPO.  As shown in IDF, this places constraints on the value of $r_\mathrm{in}$.  If $r_\mathrm{in}$ is set too close to the black hole, then the resulting precession frequency can significantly exceed observed type-C QPO frequencies.  This problem can be avoided if the flow truncates around $r_\mathrm{in} \approx 6 r_g$.  For tilted accretion flows, this requirement is actually consistent with numerical simulations in which the disk is found to truncate around $6 r_g$, independent of black hole spin \citep{Fragile09, Dexter11}.

\subsection{Initializing model tori}

\begin{figure}
  \includegraphics[width = 0.48\textwidth]{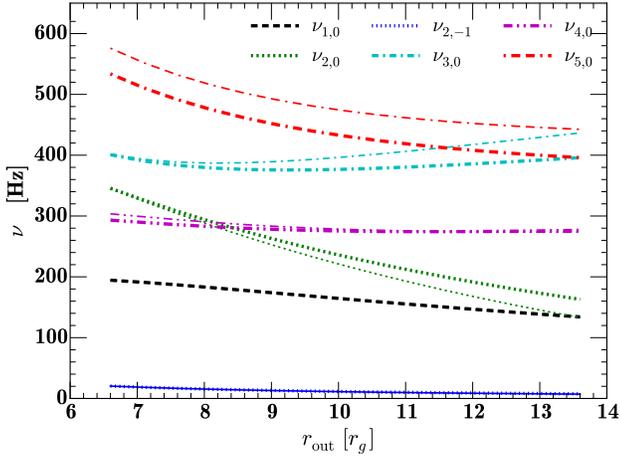}
  \caption{Frequencies of the five lowest order axisymmetric ($m = 0$) modes and
the nodal precession mode for different values of the truncation radius $r_{\rm out}$, assuming $a_* = 0.5$, $r_{\rm in} = 6 r_g$, and $n = 3$ (thick lines) or $n = 1.5$ (thin lines).}
\label{fig:modes}
\end{figure}

We are interested in comparing the lowest-order oscillation mode frequencies of tori of various sizes against the high-frequency QPOs of LMXBs.  For this purpose, it is most convenient to parametrize the tori in terms of $r_{\rm in}$ and $r_{\rm out}$, the inner and outer radial limits, respectively, rather than $r_0$ and $\beta$.  Any torus has a surface that can be defined by the equipotential 
\be
  W(r,\theta) = \ln[-u_t] = -\ln\left[\frac{(-g_{tt} - \Omega g_{tp})}{\sqrt{-gtt - 2 \Omega g_{tp} - \Omega^2 g_{pp}}}\right],
\ee
where $\Omega = -(g_{tp} + l_0 g_{tt})/(g_{pp} + l_0 g_{tp})$ is the angular velocity of the gas, which depends on the constant angular momentum $l_0$ and the metric coefficients $g_{\mu\nu}$.  In the equatorial plane ($\theta = \pi/2$), the value of the potential at the inner and outer radius is identical, $W(r_{\rm in}) = W(r_{\rm out})$.  We use this equality to solve for the value of $l_0$ for each torus. Wherever $l_0$ intersects the Keplerian angular momentum, $l_{\rm K}$, we have an equilibrium point where all forces vanish.  If $l_0 > l_{\rm ISCO}$, there are two such points, the inner saddle point (or cusp) that belongs to the largest closed torus surface and the pressure maximum point (or torus center).  We only require the location of $r_0$, as it is used to calculate the frequencies of the oscillation modes. 
In order to get the thickness of each torus, we rewrite the equipotential function so that it depends on $\beta$ and solve
\be 
  \beta^2 = \frac{2}{r_0^2 A_0^2 \Omega_0^2} \left( \ln[W(r_{\rm in})] - \ln[W_0] \right),
  \label{eq:beta}
\ee 
where $W_0 = W(r_0)$ and $A_0=u^t(r_0)$ is the redshift factor. As a reminder, we restrict ourselves to $\beta \leq 0.3$. Knowing $r_0$ and $\beta$, we can solve Equation~\ref{eq:nu_i} for each of the five oscillation modes. In Figure \ref{fig:modes}, we plot the frequencies of all five lowest order modes and the nodal precession mode over a range of values for $r_\mathrm{out}$ (with $r_\mathrm{in} = 6 r_g$ and $a_* = 0.5$). 

\subsection{3:2 frequency ratio}
\label{sec:3:2}

\begin{figure}
  \includegraphics[width = 0.48\textwidth]{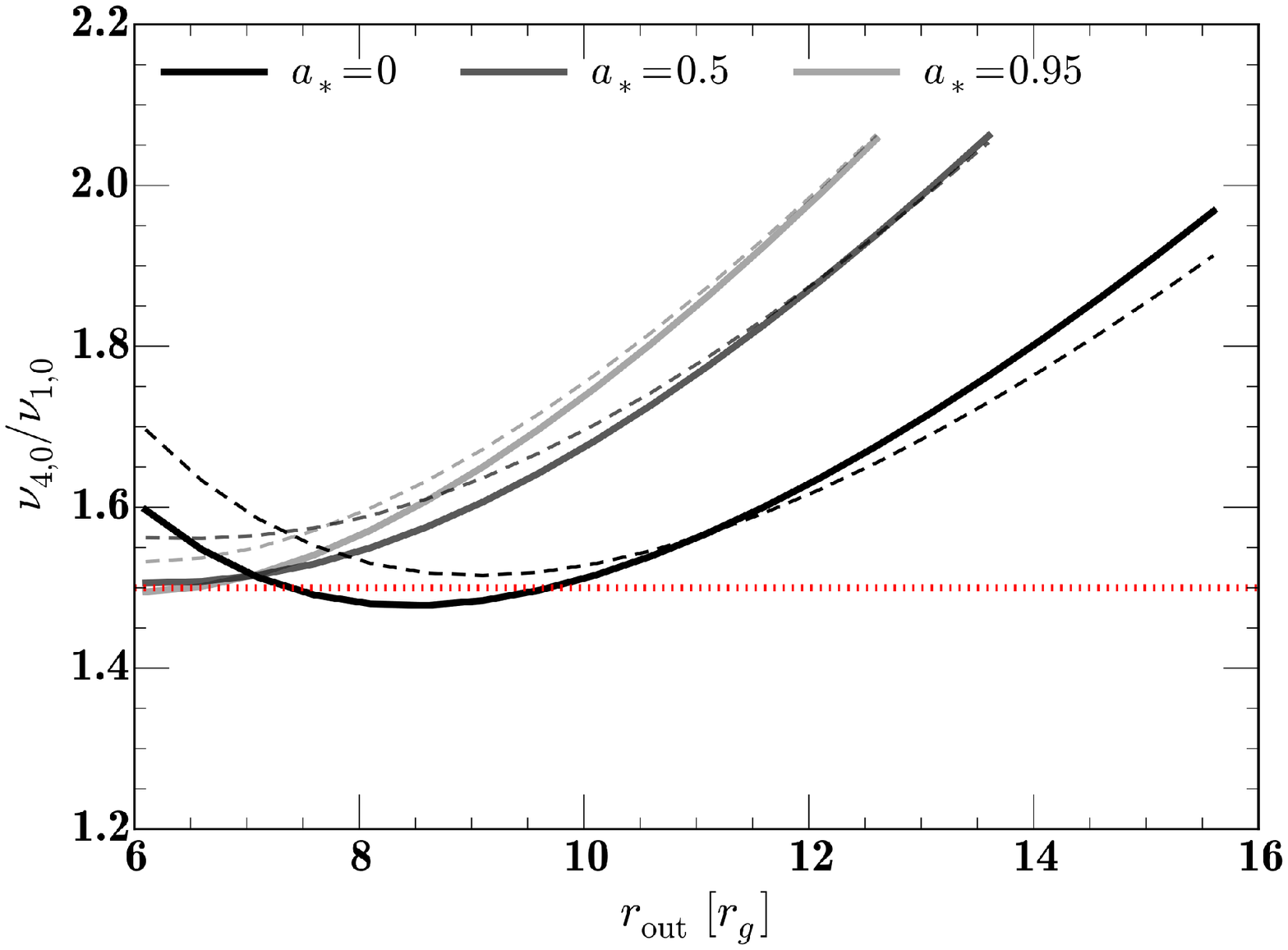}
  \includegraphics[width = 0.48\textwidth]{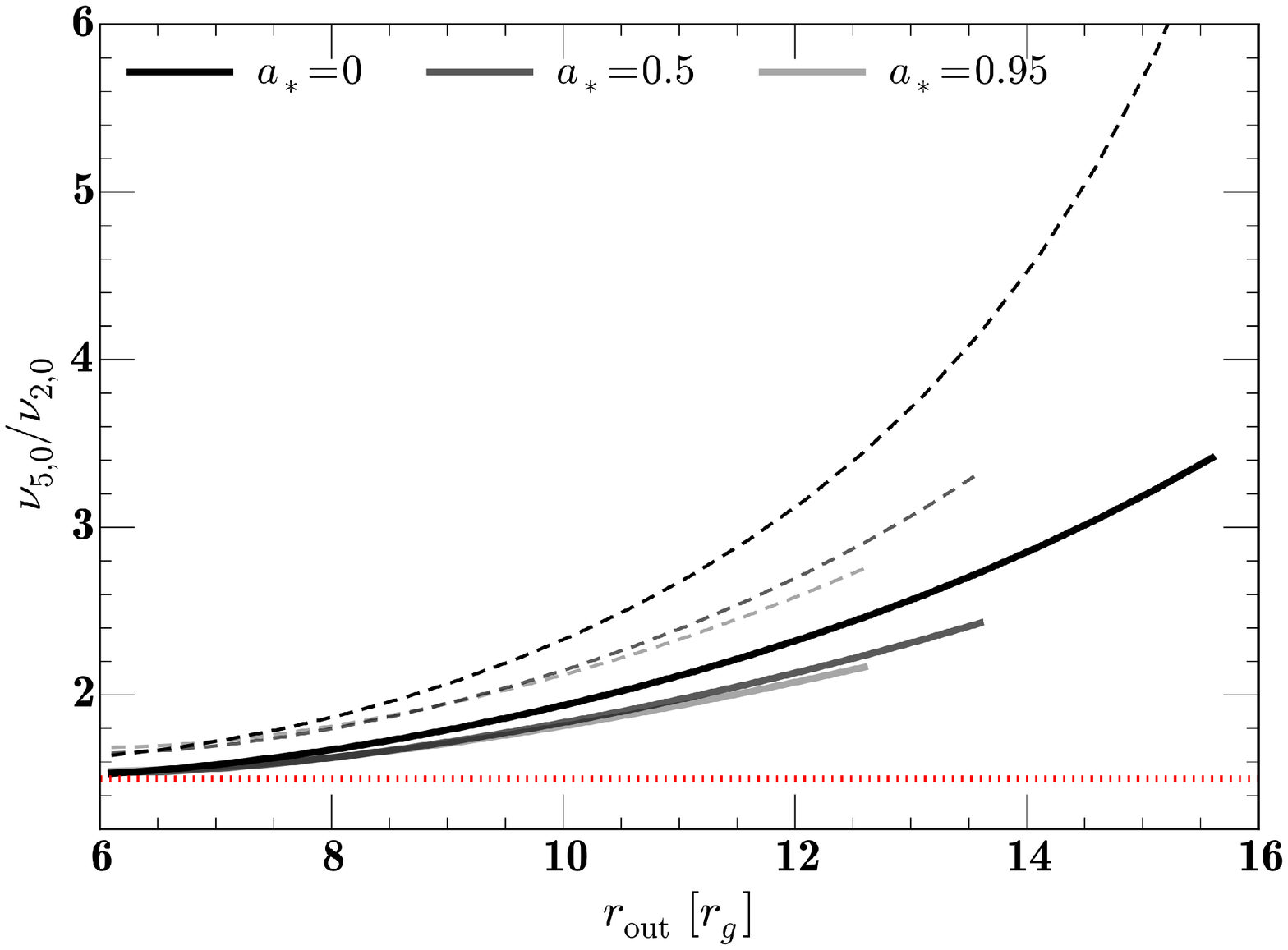}
  \includegraphics[width = 0.48\textwidth]{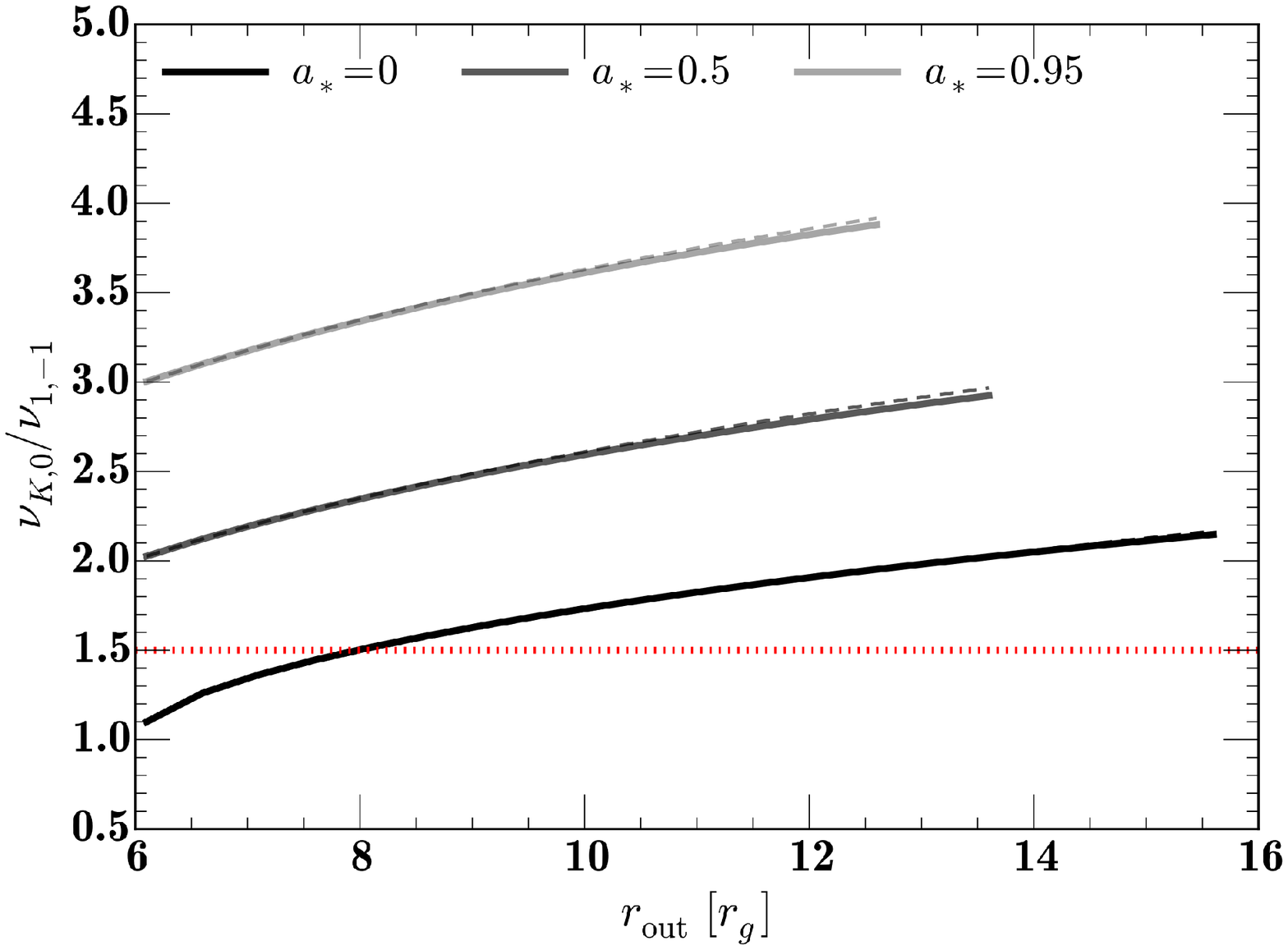}
  \caption{Ratio of the plus mode frequency to the radial epicyclic mode frequency (both $m=0$), $\nu_{4,0}/\nu_{1,0}$ ({\em top} panel); ratio of the breathing mode frequency to the vertical epicyclic mode frequency (both $m=0$), $\nu_{5,0}/\nu_{2,0}$ ({\em middle} panel); and ratio of the Keplerian frequency to the $m=-1$ radial epicyclic mode frequency, $\nu_{\rm K}/\nu_{1,-1}$ ({\em bottom} panel), for different values of the truncation radius $r_\mathrm{out}$, assuming $r_\mathrm{in} = 6 r_g$. Solid curves are for a polytropic index $n = 3$, while dashed curves are for $n = 1.5$. Each pair of curves in black, dark gray, and light gray corresponds to the spins $a_* = 0$, 0.5, and 0.95, respectively.}
\label{fig:ratio}
\end{figure}

Of the possible torus oscillation modes, \citet{Rezzolla03} showed that the
axisymmetric ($m=0$) plus ($i=4$) and radial epicyclic ($i=1$) modes have frequencies that are in a near 3:2 ratio and cover a range that may include some HFQPO pairs.  BAF did the same for the breathing ($i=5$) and vertical epicyclic ($i=2$) modes.  These same modes were analyzed in \citet{Dexter14}, although for a different black hole accretion geometry than we are considering here.  Quite recently, \citet{Torok15} suggested that the Keplerian angular frequency ($\nu_{\rm K}$) and $m=-1$, radial epicyclic ($i=1$) mode of non-slender tori be considered.  In Figure \ref{fig:ratio}, we explore how these different ratios behave in tori of different widths.  We find that the ratio of the plus mode to the radial epicyclic one ($\nu_{4,0}/\nu_{1,0}$) is close to 1.5 for a fairly wide range of values of $r_\mathrm{out}$ for both non-relativistic ($n=1.5$) and relativistic ($n=3$) gases for low to moderate spins.  The ratio of the breathing mode to the vertical epicyclic one ($\nu_{5,0}/\nu_{2,0}$), on the other hand, only approaches 1.5 in the slender limit, and even then, only for a relativistic gas ($n=3$).  The Keplerian and $m = -1$, radial epicyclic ratio passes through 1.5 {\it only} for low spins $a_* \lesssim 0.3$, but is nearly independent of the polytropic index.

\section{The Combined Model Applied to GRO J1655-40}
\label{sec:GRO}

We now apply our combined model to the case of GRO J1655-40.  As described in \citet{Motta14a}, power density spectra (PDS) of this source have shown a type-C QPO along with at least one additional feature (either a QPO or broad Lorentzian) during multiple observing epochs.  During three such observations, the type-C QPO was seen along with two HFQPOs.  To strengthen the statistics, these three observations were combined into a single PDS, for which $\nu_\mathrm{C} = 17.3\pm0.1$ Hz, $\nu_\mathrm{L} = 298\pm4$ Hz, and $\nu_\mathrm{U} = 441\pm2$ Hz.  We first attempt to fit these simultaneous QPOs with our model.  We then move on to consider all of the observations from \citet{Motta14a}.

\subsection{Fitting the Observation of Three Simultaneous QPOs}
\label{sec:GROfit}

Before proceeding, let us first recapitulate the model: we assume that the type-C QPO is associated with the global precession of a hot, thick flow, such that $\nu_\mathrm{C} = \nu_{\rm prec} = |\nu_{2,-1}$|, and that the lower and upper HFQPOs are associated with oscillation modes of this same structure, either with $\nu_\mathrm{L} = \nu_{1,0}$ and $\nu_\mathrm{U} = \nu_{4,0}$ (i.e. $m=0$ radial epicyclic and plus modes), $\nu_\mathrm{L} = \nu_{2,0}$ and $\nu_\mathrm{U} = \nu_{5,0}$ (i.e. $m=0$ vertical epicyclic and breathing modes), or $\nu_\mathrm{L} = \nu_{1,-1}$ and $\nu_\mathrm{U} = \nu_\mathrm{K}$ (i.e. the $m=-1$ radial epicyclic mode and Keplerian frequency).  The observation of three simultaneous QPOs then allows us to constrain three of the model parameters.  Formally, the model has five free parameters: the black hole mass and spin, the inner and outer radius of the hot, thick flow, and the polytropic index of the gas.  This means the model is under-constrained.  To proceed, we would like to fix the mass of the black hole in GRO J1655-40.  We choose to use $5.4 \pm 0.3 M_\odot$ \citep{Beer02}, consistent with the value used in \citet{Motta14a}.  However, our model can also be made to fit with other values, such as the $6.3 \pm 0.5 M_\odot$ of \citet{Greene01}.  We also find that our model is rather insensitive to the polytropic index, except that the breathing and vertical epicyclic modes only have a frequency ratio close to 3:2 for a relativistic gas ($n=3$), so we restrict ourselves to this value. 
  
We saw in Section~\ref{sec:3:2} that the plus and radial epicyclic modes have a frequency ratio close to 1.5 for a fairly broad range of values of $r_\mathrm{out}$, so we start by considering those two modes (i.e. $\nu_\mathrm{U} = \nu_{4,0}$ and $\nu_\mathrm{L} = \nu_{1,0}$), along with the precession frequency ($\nu_\mathrm{C} = \nu_\mathrm{2,-1}$).  However, in doing so, we are unable to fit the three simultaneous QPOs ($\nu_\mathrm{C} = 17.3\pm0.1$ Hz, $\nu_\mathrm{L} = 298\pm4$ Hz, and $\nu_\mathrm{U} = 441\pm2$ Hz) for any choices of $a_*$, $r_\mathrm{in}$, $r_\mathrm{out}$, and $n$.  The trouble is that the plus and radial epicyclic modes produce too low frequencies to match $\nu_\mathrm{U}$ and $\nu_\mathrm{L}$, unless the torus is quite slender and the spin is quite high (see Fig. 9 of \citealt{Blaes06}).  However, under these conditions, the precession frequency greatly exceeds that of the type-C QPO (see Fig. 3 of \citealt{Ingram09}).  Some sign of this can be seen in our own Figure~\ref{fig:modes}, where the plus mode only reaches $\nu_{4,0} \approx 300$ Hz, while the radial epicyclic mode never rises above $\nu_{1,0} \approx 200$ Hz.  It is worth noting that these limits only apply for axisymmetric models ($m=0$).  Non-axisymmetric modes ($m>0$) behave qualitatively similar as the axisymmetric modes but are able to reach much higher frequencies and will be studied further in future work.

For the breathing and vertical epicyclic modes (i.e. $\nu_\mathrm{U} = \nu_{5,0}$ and $\nu_\mathrm{L} = \nu_{2,0}$), we are able to find a satisfactory fit for the three simultaneous QPOs using the following model parameters (with $n=3$): $a_* = 0.63\pm0.12$, $r_\mathrm{in} = 6.5\pm0.6 r_g$, and $r_\mathrm{in}+0.2 \le r_\mathrm{out}/r_g \le r_\mathrm{in}+2.9$. Those are roughly the $1\sigma$ uncertainties, obtained by varying one model parameter at a time.  The breathing and vertical epicyclic modes are critical for constraining the size and location of the torus (note, for example, their strong dependence on $r_\mathrm{out}$ in Fig. 2), while the precession mode carries most of the dependence on spin.  Our value for the dimensionless spin is consistent with the value obtained from X-ray spectral fitting of the thermally-dominant state [$a_* = 0.65$-0.75 \citep{Shafee06}], but not with the higher values obtained from Fe-line reflection fitting [$a_* = 0.94$-0.98 \citep{Miller09}; $a_* = 0.9$ \citep{Reis09}].  Our value for $r_\mathrm{in}$ places the inner edge of the torus outside of $r_\mathrm{ISCO}$.  This may be consistent, if the flow is tilted, as simulations of tilted, hot, thick flows exhibit a disk truncation radius of $r_\mathrm{in} \approx 6 r_g$, regardless of black hole spin \citep{Fragile09,Dexter11}.

We can also fit the GRO J1655-40 data using the Keplerian and $m=-1$, radial epicyclic modes (i.e. $\nu_\mathrm{U} = \nu_{\rm K}$ and $\nu_\mathrm{L} = \nu_{1,-1}$).  In this case, the model requires a much lower black hole spin: $a_* = 0.29 \pm 0.03$.  It also requires the torus to lie closer to the black hole: $r_\mathrm{in} = 5.3 \pm 0.3$; with $r_\mathrm{in}+0.3 \le r_\mathrm{out}/r_g \le r_\mathrm{in}+1.5$.  Thus, although this set of modes can fit the simultaneous observations of the three QPOs, it does so with a very different torus configuration than the breathing/vertical epicyclic mode combination and a black hole spin that is inconsistent with most other estimates.  Future observations may be able to use these distinctions to favor one set of modes over the other.

\subsection{Extending the Model to Lower Frequencies}

\begin{figure}
  \includegraphics[width = 0.48\textwidth]{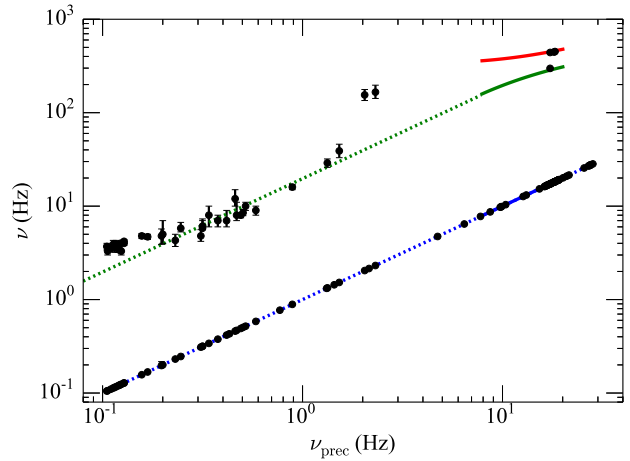}
  \caption{Breathing mode ({\em top, red}), vertical epicyclic mode ({\em middle, green}), and precession frequency ({\em bottom, blue}) plotted against the precession frequency for a model with $M_\mathrm{BH} = 5.4 M_\odot$, $a_* = 0.63$, $r_\mathrm{in} = 6.5 r_g$, and $n = 3$.  The outer radius, $r_\mathrm{out}$, is allowed to vary, thus producing the range of frequencies.  The torus modes can only be accurately estimated for slightly non-slender tori ({\em solid lines}).  We extrapolate the vertical epicyclic, $\nu_2$, and precession frequencies to lower frequencies ({\em dotted line}) for illustration. QPO data from \citet{Motta14a} are included, with the QPO frequencies plotted against the type-C QPO.}
\label{fig:qpos}
\end{figure}

In the simplest implementation of our model, we assume that all of its parameters remain fixed for a given source, except for the truncation radius, $r_\mathrm{out}$, between the cold, thin disk and the hot, thick flow.  As a given outburst proceeds, all of the evolution in the QPO frequencies is then controlled directly by the change of $r_\mathrm{out}$.  To explore how this works, we now fix all of the parameters for GRO J1655-40, except $r_\mathrm{out}$, using the results of Section~\ref{sec:GROfit} for the breathing/vertical epicyclic mode pair (i.e. $M_\mathrm{BH} = 5.4 M_\odot$, $a_* = 0.63$, $r_\mathrm{in} = 6.5 r_g$, and $n = 3$).  As a reminder, because Equation~\ref{eq:nu_i} is based on Taylor expansions about the slender torus solution, we can not extend our model to arbitrarily large tori.  For this reason, we are unable to directly apply our model to the broad Lorentzian components seen at $\nu < 40$ Hz.  This would require a more general solution of (non-slender) torus oscillation modes or else direct numerical simulations.  Nevertheless, we show in Figure~\ref{fig:qpos} that a simple extrapolation of the vertical epicyclic mode ($\nu_2$) to lower frequencies passes through or quite close to most of the broad Lorentzian features.  We note that a similar extrapolation of the $m=-1$, radial epicyclic mode does {\em not} pass through the broad Lorentzian features.  Since the type-C QPO ($\nu_\mathrm{C}$) and precession frequencies ($\nu_\mathrm{prec}$) are plotted against themselves in Figure~\ref{fig:qpos}, they, of course, lie exactly on one another.  It is important to emphasize that only a single model parameter is varied to produce the full range of QPO frequencies seen in this figure.  It is remarkable that such a simple model may be able to explain QPO observations spanning more than two decades in total range.  Only two of the broad Lorentzians, the ones at $\nu = 155$ and 166 Hz, are left unexplained by our model.

\section{Discussion and Conclusions}
\label{sec:discuss}

In this paper we have presented a relatively simple model for three commonly observed QPOs in LMXBs.  In this picture, the upper and lower HFQPOs are associated with oscillation modes of a hot, thick accretion flow, while the type-C QPO is associated with its global precession.  By applying this model to the three simultaneous QPOs observed in GRO J1655-40, we are able to narrow down the possible pairs of oscillation modes that fit the data to either the breathing and vertical epicyclic modes or the Keplerian orbital frequency and the $m=-1$, radial epicyclic mode.  The plus and ($m=0$) radial epicyclic modes appear to be ruled out for this source provided the type-C QPO is owing to the precession of the same structure that is oscillating. 

The model not only fits the simultaneous appearance of all three QPOs, but a simple extrapolation appears to connect to timing features found at lower frequencies. (This only applies to the precession and vertical epicyclic modes; we discuss the different behavior of the breathing mode below.)  In this picture, the lower HFQPO and type-C QPO frequencies are expected to drift during an outburst as the truncation radius of the thin disk moves inward, forcing the hot, thick flow to become more compact.  
All of this rich behavior comes from easily understood physics associated with hot, thick flows.  Further, the presence of such a hot, thick flow is predicted by the truncated disk model of the low, hard state, in which these QPOs appear.  

In this first paper, we have not addressed the radiation mechanisms of these QPOs.  We plan to consider this issue more carefully in future work.  For now, we make only a few qualitative statements.  First, there would seem to be plenty of power available.  If one makes the simple assumption that the hot thick flow dissipates some significant fraction, $f$, of the gravitational potential energy still available inside of the truncation radius, then the luminosity directly associated with it would be
\begin{equation}
L_\mathrm{thick} \sim f G \dot{M} \left(\frac{1}{r_\mathrm{in}} - \frac{1}{r_\mathrm{out}}\right) ~.
\end{equation}
Additionally, the common association of HFQPOs with the steep power law state \citep[e.g.][]{Remillard06} might be explained by something like the \citet{Dexter14} model, where the QPOs are coming from a region that is effectively optically thin, yet radiatively {\em efficient} (not inefficient).  

To explain how the oscillation modes might imprint themselves into the observed light curves, we refer to results already in the literature.  \citet{Ingram12} showed that a precessing thick flow (equivalent to our $m=-1$ vertical epicyclic mode) can modulate a light curve by changing both the projected solid angle and the seed photon interception geometry \citep[more detailed analysis is presented in][]{Veledina13}.  We would expect the ($m=0$) vertical epicyclic mode to modulate the light curve in much the same way.  The breathing mode, however, may behave in a fundamentally different way.   Unlike the other two modes, which are incompressible, the breathing mode is compressible.  As such, it most likely would manifest itself through modulations of the density and temperature of the torus \citep[see, e.g.,][]{Mishra15}.  At this time, we do not have an explanation for why this mode does not appear to connect to any features at lower frequencies.  This, too, is something we hope to investigate in future work.

In this paper, we focused the application of our model on GRO J1655-40.  However, that is not the only source to show multiple HFQPOs along with a type-C QPO.  XTE J1550-564, for instance, has been observed to show very similar frequency correspondence \citep{Motta14b}, and H1743-322 has shown simultaneous HFQPOs with a 3:2 frequency ratio and lower frequency QPOs \citep{Homan05}.  Recently, similar claims have also been made for two intermediate mass black hole candidates: M82 X-1 \citep{Pasham14} and NGC 1313 X-1 \citep{Pasham15}.  We plan to test our model against these sources in future work.  

\section{Acknowledgements}

We thank M. Abramowicz, J. Dexter, C. Done, A. Ingram, and S. Motta for helpful discussions and comments. We also thank the anonymous referee for helpful suggestions for improvement. This research was supported by the National Science Foundation grant NSF AST-1211230.  P.C.F. and O.B. thank the International Space Science Institute, where part of this work was carried out, for their hospitality.

\end{document}